\documentclass{aa}  
\usepackage{color}
\usepackage{graphicx}
\usepackage{lscape}
\usepackage{natbib}
\usepackage[varg]{txfonts}
\usepackage{url}
\usepackage{xspace}
\bibpunct{(}{)}{;}{a}{}{,}

\newcommand{\Teff}{$T\mathrm{\hspace*{-0.4ex}_{eff}}$\,}
\newcommand{\logg}{$\log\,g$\hspace*{0.5ex}}
\def\egb{\object{EGB\,6}}
\def\tol{\object{Tol\,26}}
\begin{document}

\title{The hot white dwarf in the peculiar binary nucleus of the planetary
  nebula EGB\,6\thanks{Based on observations made with the
    NASA-CNES-CSA Far Ultraviolet Spectroscopic Explorer and the
    NASA/ESA Hubble Space Telescope, obtained at the Space Telescope
    Science Institute, which is operated by the Association of
    Universities for Research in Astronomy, Inc., under NASA contract
    NAS5-26666.}}

\author{K\@. Werner\inst{1} \and T\@. Rauch\inst{1} \and J\@. W\@. Kruk\inst{2}}

\institute{Institute for Astronomy and Astrophysics, Kepler Center for
  Astro and Particle Physics, Eberhard Karls University, Sand~1, 72076
  T\"ubingen, Germany\\ \email{werner@astro.uni-tuebingen.de} 
\and
  NASA Goddard Space Flight Center, Greenbelt, MD\,20771, USA }

\date{Received 10 April 2018 / Accepted 14 May 2018}


\abstract{\egb\ is an extended, faint old planetary nebula (PN) with
  an enigmatic nucleus. The central star (PG\,0950$+$139) is a hot
  DAOZ-type white dwarf (WD). An unresolved, compact emission knot was
  discovered to be located $0\farcs166$ away from the WD and it was
  shown to be centered around a dust-enshrouded low-luminosity
  star. It was argued that the dust disk and evaporated gas
  (photoionized by the hot WD) around the companion are remnants of a
  disk formed by wind material captured from the WD progenitor when it
  was an asymptotic giant branch (AGB) star. In this paper, we assess the
  hot WD to determine its atmospheric and stellar parameters. We
  performed a model-atmosphere analysis of ultraviolet (UV) and
  optical spectra. We found \Teff = $105\,000 \pm 5000$\,K, \logg =
  $7.4 \pm 0.4$, and a solar helium abundance (He = $0.25\pm 0.1$,
  mass fraction). We measured the abundances of ten more species (C,
  N, O, F, Si, P, S, Ar, Fe, Ni) and found essentially solar abundance
  values, indicating that radiation-driven wind mass-loss, with a
  theoretical rate of $\log$($\dot{M}/M_\sun$/yr)
  $=-11.0^{+1.1}_{-0.8}$ prevents the gravitational separation of
  elements in the photosphere. The WD has a mass of
  $M/M_\sun=0.58^{+0.12}_{-0.04}$ and its post-AGB age ($\log$($t_{\rm
    evol}$/yr = $3.60^{+1.26}_{-0.09}$)) is compatible with the PN
  kinematical age of $\log$($t_{\rm PN}$/yr = $4.2$). In addition, we
  examined the UV spectrum of the hot nucleus of a similar object with
  a compact emission region, \tol\ (PN\,G298.0$+$34.8), and found that
  it is a slightly cooler DAOZ WD (\Teff $\approx$\,85\,000\,K), but
  this WD shows signatures of gravitational settling of heavy
  elements.}


\maketitle
%

\begin{figure*}[t]
 \centering  \includegraphics[width=1.0\textwidth]{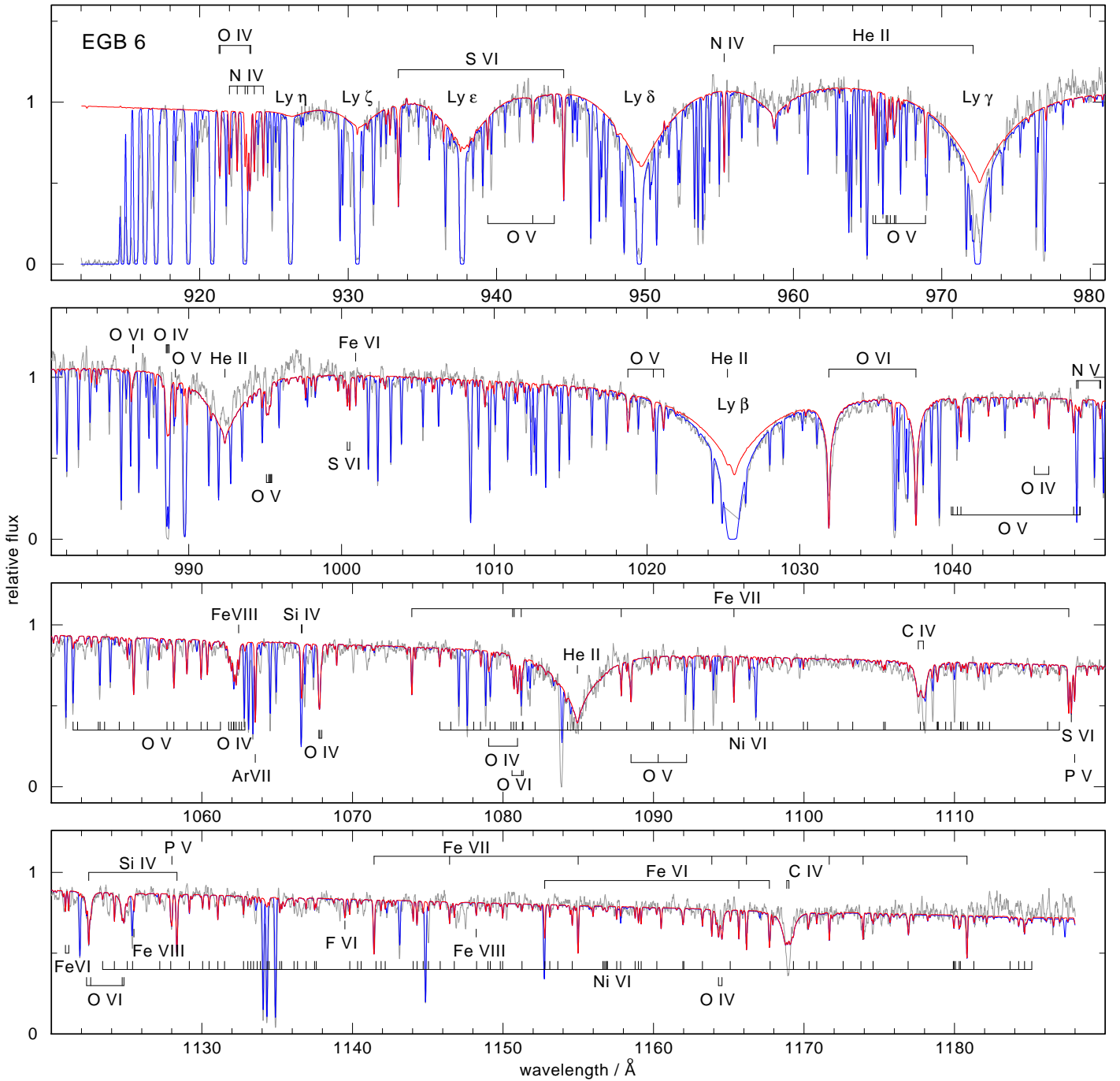}
  \caption{FUSE spectrum of the white dwarf in \egb\ (gray) compared to a
    photospheric model spectrum (red; \Teff = 105\,000\,K, \logg
    = 7.4) with the  adopted parameters  (see
    Table\,\ref{tab:results}). The same model attenuated by
    interstellar lines is plotted in blue. Prominent
    photospheric lines are identified.}
\label{fig:egb6_fuse}
\end{figure*}

\begin{figure*}[t]
 \centering  \includegraphics[width=1.0\textwidth]{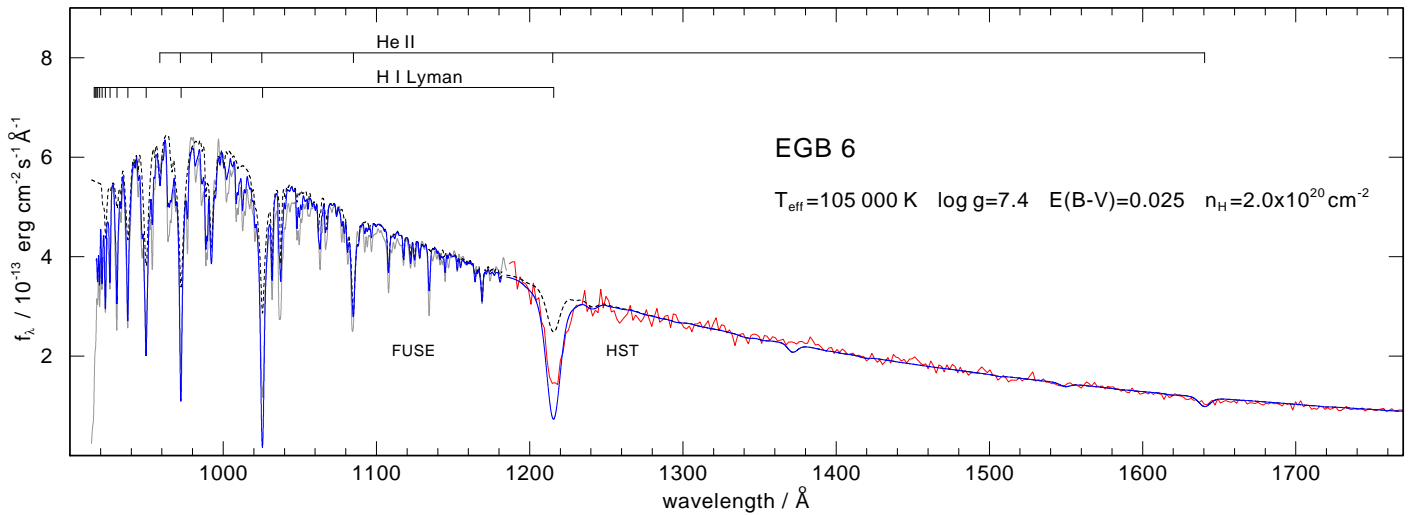}
  \caption{FUSE spectrum (gray) and HST/FOS spectrum (red) of the
white dwarf in \egb\ compared to a photospheric model
    spectrum (blue; \Teff = 105\,000\,K, \logg = 7.4) with the
     adopted parameters (see     Table\,\ref{tab:results}). A reddening of $E(B-V)=0.025$ was
    applied to the model to fit the observed continuum shape. The
    model is attenuated by interstellar Lyman line absorption (column
    density $n_{\rm H}=2\times 10^{20}$\,cm$^{-2}$). The black dashed
    line is the unattenuated model. For clarity, the FUSE spectrum and
    the models in the respective wavelength range were smoothed with a
    Gaussian with 1\,\AA\ FWHM. The model in the HST/FOS range was
    convolved with a 7\,\AA\ Gaussian.}
\label{fig:overview_uv}
\end{figure*}

\begin{figure}[t]
 \centering
 \includegraphics[width=0.6\columnwidth]{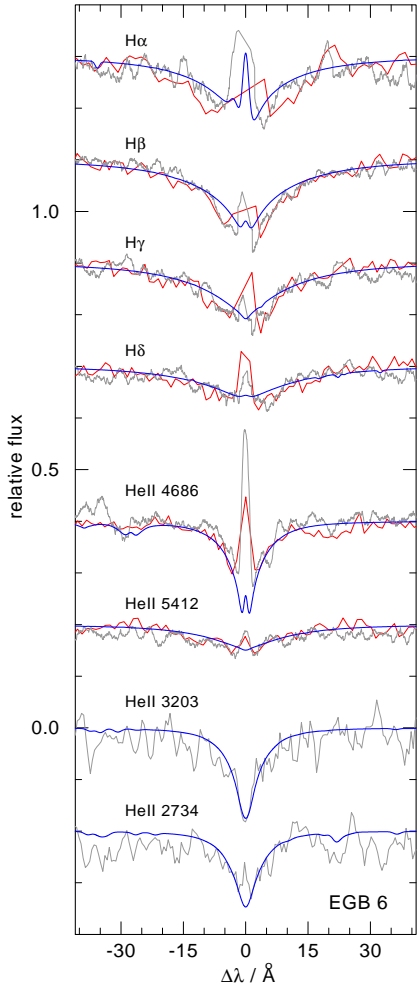}
  \caption{Observed Balmer and \ion{He}{ii} lines of the white dwarf in
    \egb\ (red: SDSS spectrum, gray: SPY spectrum; two lowest lines:
    HST spectrum) compared to a model spectrum (blue graph; \Teff =
    105\,000\,K, \logg = 7.4) with the  adopted parameters (see Table\,\ref{tab:results}). Nebular emission lines in the
    photospheric Balmer line cores are truncated for clarity.}
\label{fig:optical}
\end{figure}

\begin{figure*}[t]
 \centering
 \includegraphics[width=1.0\textwidth]{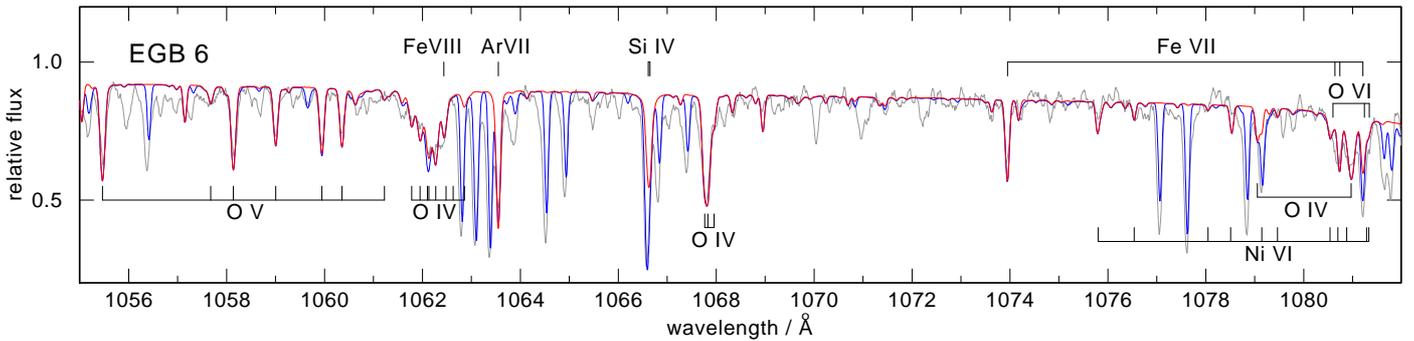}
  \caption{Detail of the FUSE spectrum and model for the white dwarf in
    \egb\ from Fig.\,\ref{fig:egb6_fuse}.}
\label{fig:egb6_fuse_detail}
\end{figure*}

\begin{figure}[t]
 \centering  \includegraphics[width=0.95\columnwidth]{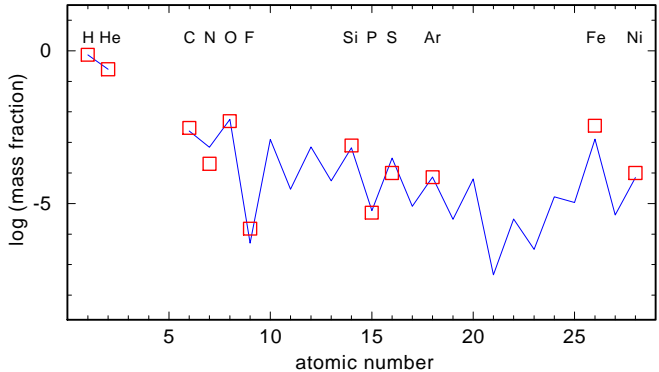}
  \caption{Element abundances measured in the white dwarf in \egb\ (red
    squares; see also Table\,\ref{tab:results}). The blue line indicates solar
    abundances.}\label{fig:abu}
\end{figure}

\begin{figure*}[t]
 \centering  \includegraphics[width=1.0\textwidth]{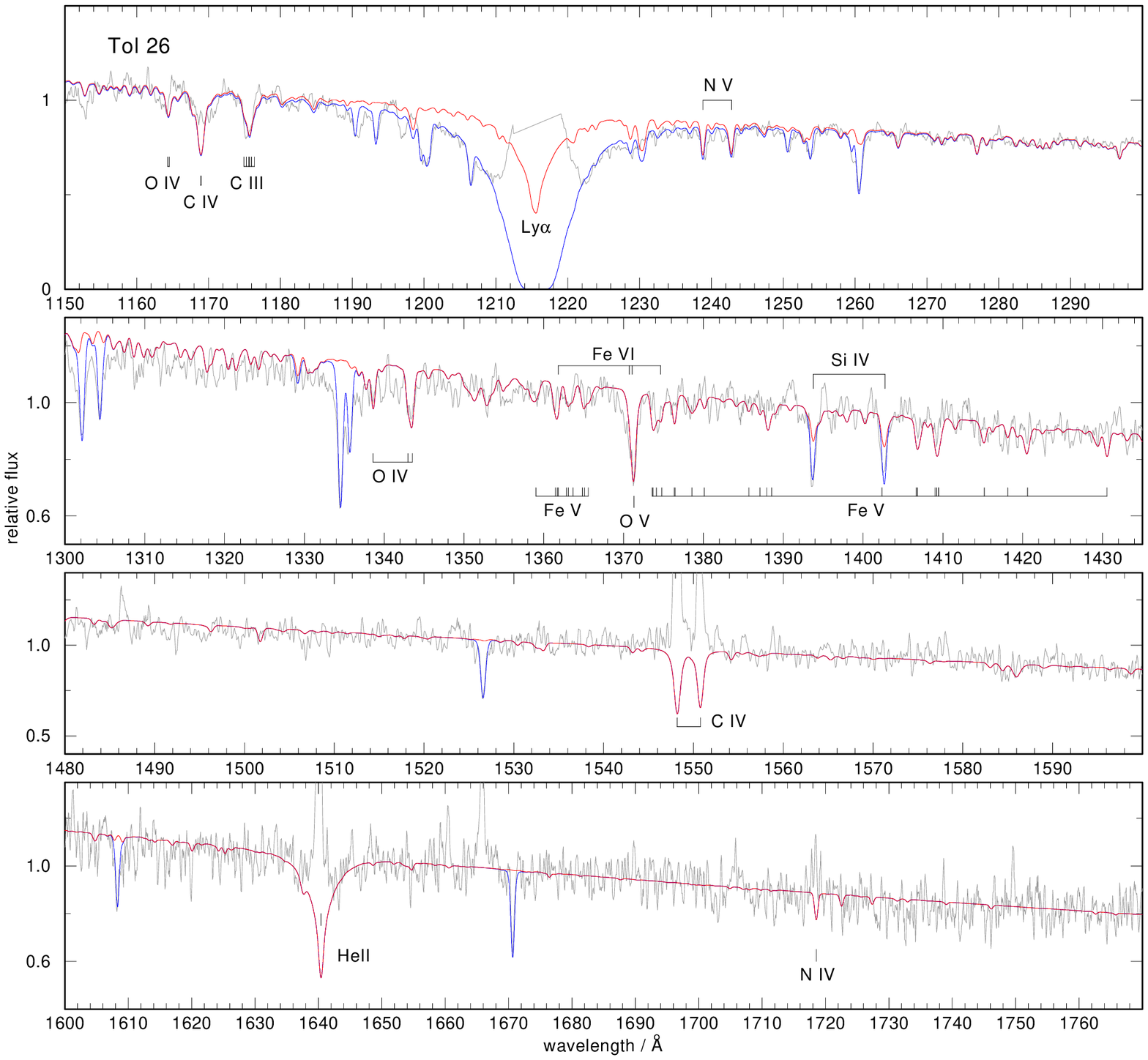}
  \caption{HST/COS spectrum of the white dwarf in \tol\ (gray) compared
    to a photospheric model spectrum (red; \Teff = 85\,000\,K,
    \logg = 7.4) with abundances as given in
    Sect.\,\ref{sect:tol26}. The blue graph is the photospheric model
    attenuated by interstellar lines. Prominent photospheric lines are
    identified. Absorption profiles of the \ion{C}{iv} resonance
    doublet and \ion{He}{ii} $\lambda$1640\,\AA\ (except for the far
    wings) are masked by nebular emission lines.}
\label{fig:tol26}
\end{figure*}

\section{Introduction}
\label{sect:intro}

\egb\ (PN\,G221.5$+$46.3) is a faint old planetary nebula (PN) with a
mysterious nucleus. The PN was discovered by
\citet{1984PASP...96..283E} and has an angular extension of $11\arcmin
\times 13\arcmin$. At a nominal distance of $\approx$\,725\,pc, the linear
diameter of the PN is $\approx$\,1\,pc. The central star (PG\,0950$+$139)
is a very hot white dwarf (WD) that was independently discovered by
\citet{1984PASP...96..283E} and  the Palomar Green (PG) survey
\citep{1986ApJS...61..305G}. Bright nebula emission lines superposed
on the WD spectra were shown not to stem from the old, low-surface
brightness PN, but from a relatively dense \citep[$n_e =2.2 \times
  10^6$\,cm$^{-3}$,][]{1989ApJ...347..910D}, unresolved compact
emission knot (CEK) that was thought to be centered on the WD. An
infrared excess was discovered by \citet{1991ASPC...14..161Z} and
attributed to the presence of an M-type stellar companion. It came as a
big surprise when imaging and grism spectroscopy obtained with the
\emph{Hubble} Space Telescope (HST) revealed that the CEK is a
point-like source at the location of the companion, $0\farcs166$
away from the WD \citep{2013ApJ...769...32L}, corresponding to a
projected linear separation of $\approx$\,118\,AU. The origin of the
material making up the CEK is unknown. It was suggested that it is the
remnant of a disk containing material captured by the companion star
from the WD progenitor's wind during its asymptotic giant branch (AGB)
stage, and that this disk is surrounded by gas (the CEK) photoionized
by the hot WD. The nature of the companion is in fact unknown because
it is enshrouded by dust \citep{2016ApJ...826..139B}. It must be a
low-luminosity object, a dM star or even another WD. For a
detailed summary of our current knowledge about this system, we refer
to the exhaustive paper by \citet{2016ApJ...826..139B} and references
therein. 

In the present paper, we investigate in detail the hot central star of
\egb.  It was classified as a hot DA white dwarf by
\citet{1986ApJ...308..176F}. Later on, by the detection of ionized helium,
\citet{1989ApJ...346..251L} established the star as a DAO white
dwarf, and after the detection of metal lines in the ultraviolet
\citep{2010ApJ...720..581G} the star was assigned to the DAOZ class.

The analysis of the optical spectrum of the WD is hampered by strong
emission lines from the CEK.  Using pure-hydrogen local thermodynamic
equilibrium (LTE) model atmospheres, \citet{1989ApJ...346..251L}
derived \Teff = $70\,000\pm 7000$\,K and \logg = $7.5\pm 0.25$ from a
fit to the H$\gamma$ line only. From new observations, and with
pure hydrogen non-LTE models for Balmer line fits,
\cite{2005ApJS..156...47L} found \Teff = $108\,390 \pm 16\,687$\,K and
\logg = $7.39\pm0.38$.

Using non-LTE model atmospheres, \citet{2010ApJ...720..581G} derived
\Teff = $93\,230 \pm 20\,425$\,K and \logg = $7.36 \pm 0.79$,
emphasizing increased uncertainties in the measured parameters due to
the exclusion of the Balmer line cores from the fitting procedure. For
the first time, they measured the helium abundance
(He/H=0.1$^{+0.7}_{-0.09}$, by mass) which, at face value, is lower
than the solar abundance (He/H=0.34), but highly uncertain. In
  their hydrogen plus helium models, in order to compute a more
  realistic atmospheric structure, they included the elements C, N, and O  in solar
  amounts, as prescribed by the analysis of
  \citet{1996ApJ...457L..39W}, who outlined the solution to the
  Balmer line problem. This measure was motivated by the fact that
they identified metal lines in Far Ultraviolet Spectroscopic Explorer
(FUSE) spectra; however, abundance determinations were not performed.

Here, we present a new spectral analysis of the WD. Our primary aim was
to determine the metal abundances from the ultraviolet (UV) spectra taken
with FUSE and to improve the measurement of \Teff, \logg, and He
abundance employing, in addition, previously unused optical spectra. It
was expected that the results would contribute one piece to resolve the
puzzle of this extraordinary PN nucleus.\footnote{To pick up the
  thread from \citet{2016ApJ...826..139B}, a fitting quotation of the
  fictional Sherlock Holmes might be: ``It is, of course, a trifle,
  but there is nothing so important as trifles.''
  \citep{conandoyle}. }

It has been claimed that \egb\ is the prototype of a small class of PNe
with central compact emission regions
\citep{2010PASA...27..129F}. Related to them could be
\tol\ \citep[PN\,G298.0$+$34.8, ][]{1981PASP...93...93H}, whose
central star was hitherto unexplored. From the nebula analysis,
\Teff$>50\,000$\,K was concluded. Here we present the investigation of
an archival UV spectrum obtained with the Cosmic Origins Spectrograph
(COS) aboard HST, to check for the possible similarity between  the hot central stars of
\egb\ and \tol .

\section{Observations}
\label{sect:observations}

For our analysis of \egb, we use UV spectra which were
observed with the International Ultraviolet Explorer (IUE), HST, and
FUSE. Optical spectra were obtained by the Sloan Digital Sky Survey
(SDSS) and the Supernovae Type Ia Progenitor Survey
\citep[SPY,][]{2003Msngr.112...25N}. For \tol, we used the HST/COS spectrum.

\subsection{Ultraviolet spectra}

\egb\ was observed  by FUSE twice in 2000 and we retrieved the spectra
from the MAST archive. The first observation (A0340101) had imprecise
target coordinates, so the star was on the edge of the slit in two
channels and completely missed the other two. The second observation
(A0341101) was good in all channels, so we scaled the usable data from
the first observation to match the second. The reduction procedure was
the same as that described in \citet{2016A&A...593A.104W}. The
resulting spectrum is shown in Fig.\,\ref{fig:egb6_fuse}. The
resolving power is about 20\,000. We smoothed the spectrum by a
0.05\,\AA\ wide boxcar and, accordingly, our synthetic spectra were
convolved with a Gaussian with FWHM=0.07\,\AA.

The hydrogen Lyman lines are blended with the respective \ion{He}{ii}
line series. From the latter, three isolated lines are visible, with
decreasing strength, at 1085\,\AA, 992\,\AA, and
959\,\AA\ (transitions between levels with principal quantum numbers
$n=2\rightarrow 5, 6, 7$). The FUSE spectrum was published by
\citet{2010ApJ...720..581G} and  photospheric lines of seven metals
were identified (C, N, O, Si, P, S, Fe); however, it was not subject
to a quantitative analysis. We identified many more lines of these
elements plus three more species (F, Ar, and Ni; see
Sect.\,\ref{sect:metals}). 

The FUSE observations exhibit many absorption lines of the ISM. To
unambiguously identify stellar lines, we employed the program OWENS
\citep{2002ApJS..140...67L,2002P&SS...50.1169H,2003ApJ...599..297H}. It
has the capacity to consider several, individual ISM clouds with their
own radial and turbulent velocity, temperature, chemical composition,
and respective column densities. We identified and modeled lines of
\ion{H}{i}, \ion{D}{i}, H$_2$ ($J$ = 0--4), HD ($J$ = 0),
\ion{C}{ii--iii}, \ion{C}{ii}$^*$, \ion{N}{i--iii}, \ion{O}{i},
\ion{Si}{ii}, \ion{P}{ii}, \ion{S}{iii}, \ion{Ar}{i}, and
\ion{Fe}{ii}.

The star was observed in 1993 with the Faint Object Spectrograph (FOS)
aboard HST. Two spectra were taken with gratings G160L
(1150--2500\,\AA, resolution $\approx$\,7\,\AA) and G270H (2220--3300\,\AA,
resolution $\approx$\,2\,\AA). Because of the low resolution of the G160L
spectrum, even the strongest spectral lines predicted by our models
(\ion{He}{ii} $\lambda$1640, \ion{C}{iv} $\lambda$1550, \ion{N}{v}
$\lambda$1240, \ion{O}{v} $\lambda$1371) are not detected beyond
doubt. In the G270H spectrum, three members of the \ion{He}{ii} Fowler
series can be seen ($n=3\rightarrow 5, 6, 7$).

Three observations during 1987--1989 were performed with IUE (SWP
camera, resolution 5\,\AA). The co-added spectrum was taken from the
database created by \citet{2003ApJS..147..145H}. Its flux level
matches the HST/FOS data. At best, \ion{He}{ii} $\lambda$1640 and
\ion{C}{iv} $\lambda$1550 are barely detectable.

The UV spectra up to $\lambda = 1770$\,\AA\ are shown in
Fig.\,\ref{fig:overview_uv}. We fit the continuum shape with our final
photospheric model assuming interstellar reddening $E(B-V)=0.025\pm
0.005$, in good agreement with the value derived by
\citet[][$E(B-V)=0.02$]{2016ApJ...826..139B} and confirming the
conclusion that the hot central star suffers no reddening within the
system because the total reddening in the direction of \egb\ is about
$E(B-V)=0.027$ \citep{2011ApJ...737..103S}. From the Lyman lines, we
derive an interstellar neutral hydrogen column density of
$\log$($n_{\rm H}/$cm$^2$)$ = 20.3\pm 0.4$. This mainly follows from
fitting the Ly$\alpha$ profile because only for this line is the
equivalent width  dominated by the interstellar medium (ISM) and not
by the photospheric contribution.

A HST/COS spectrum of the central star of \tol\ was recorded in 2014
with grating G140L. It has a useful wavelength range of
$\approx$\,1160--2000\,\AA\ and a spectral resolution of
$\approx$\,0.6\,\AA. It is  discussed in detail in
Sect.\,\ref{sect:tol26}. The spectrum was smoothed with a
0.1\,\AA\ wide boxcar and the models were folded with
0.6\,\AA\ Gaussians.

\subsection{Optical spectra}

The archival SDSS spectrum of the \egb\ nucleus covers the wavelength
range 3600--10\,300\,\AA\ with a resolution of
$\approx$\,2.5\,\AA. The spectrum is a co-addition of four single
observations. We rectified it for our fitting with normalized
synthetic line profiles (Fig.\,\ref{fig:optical}). The only
photospheric lines that can be identified are from H and
\ion{He}{ii}. The cores of the Balmer lines and \ion{He}{ii}
$\lambda$4686 (and perhaps also \ion{He}{ii} $\lambda$5412) are
contaminated with emission lines from the CEK around the companion
\citep{2013ApJ...769...32L}, hampering the determination of
atmospheric parameters. 

To mitigate this problem, we used SPY spectra, which have a higher
resolution ($\approx$\,0.1\,\AA), allowing us to trace the photospheric
line profiles closer toward the line core, because the nebular
emission lines are rather narrow. However, the signal-to-noise ratio
(S/N) of these spectra is relatively poor. In total, three SPY spectra
(each consisting of three segments covering the ranges
3400--4520\,\AA, 4620--5600\,\AA, and 5680--6640\,\AA) were used and
rectified. We abandoned co-addition because of the very different S/N
of the single spectra in different wavelength regions. In the SPY
spectra, too, no metal lines are detectable. 

The \ion{He}{ii} $\lambda$3203 line, covered by the HST/FOS G270H
spectrum, is obviously not affected by nebula emission. The evaluated
optical and near-UV  H and \ion{He}{ii} lines are shown in
Fig.\,\ref{fig:optical}.

\section{Spectral analysis}
\label{sect:results}

We used the T\"ubingen Model-Atmosphere Package
(TMAP\footnote{\url{http://astro.uni-tuebingen.de/~TMAP}}) to compute
non-LTE, plane-parallel, line-blanketed atmosphere models in radiative
and hydrostatic equilibrium
\citep{1999JCoAM.109...65W,2003ASPC..288...31W,tmap2012}. The models
include H, He, C, N, O, Si, P, S, Fe, and Ni. The employed model atoms
are  described in detail by
\citet{2018A&A...609A.107W}. In addition, we performed line formation
iterations (i.e., keeping  the atmospheric structure fixed) for argon
and fluorine using the model atoms presented in
\citet{2015A&A...582A..94W}.

We first turn to the WD in
\egb\ (Sects.\,\ref{sect:teff}--\ref{sect:para}) and then to the
nucleus in \tol\ (Sect.\,\ref{sect:tol26}).

\subsection{Effective temperature, surface gravity, He abundance}
\label{sect:teff}

We started our spectrum fitting procedure for \egb\ from a model with
temperature, gravity, and helium abundance close to the values derived
by \citet{2010ApJ...720..581G}: \Teff = 93\,000\,K, \logg = 7.4,  H
= 0.9 and He = 0.09 (mass fractions), and assuming solar abundances for
the metals. Keeping fixed \logg\ and the H/He ratio, we then adjusted
iteratively \Teff\ and the metal abundances to achieve a good fit to the
metal lines in the FUSE spectrum. From the relative strengths of lines
from different ionization stages of nitrogen, oxygen, and iron (see
Sect.\,\ref{sect:metals}), we realized that the effective temperature
must be higher, namely 105\,000\,K, within a narrow margin of about
5000\,K. Then the possible range of the surface gravity was checked
with the optical H and He lines, confirming the initial value within
\logg = $7.4 \pm 0.4$. The UV lines of H and He are compatible with
this result but they are less sensitive to \logg. Then the He
abundance was refined by using the optical and UV spectra and we found
a better fit with a higher He/H ratio, namely a solar value (H = 0.74
and He = 0.25). A model computed with \logg = 8.0 shows H and He lines
that are significantly too broad.

To investigate the effects of the uncertainty in \logg\ on the other
atmospheric parameters, we repeated the fitting procedure with \logg =
7.0 models. The lower gravity leads to lower particle densities and,
according to the Saha equation, metal ionization balances are shifted
to higher stages. To compensate for this, \Teff\ must be reduced to
retain good line fits, namely down to 100\,000\,K. Metal abundances
are hardly affected by this change in \Teff\ and \logg, but the helium
abundance must be reduced in the model. Otherwise, the \ion{He}{ii}
lines become too strong, in particular those at 959\,\AA, 4686\,\AA,
and 5412\,\AA, and a detectable $\lambda$4542 line is predicted that
is not seen in the observations. We arrive at He = 0.18 (and thus H =
0.81). From these experiments we conclude \Teff = $105\,000 \pm
5000$\,K, \logg = $7.4\pm 0.4$, and He = $0.25 \pm 0.1$. The error in
the individual metal abundances is estimated not to exceed $\pm
0.5$\,dex.

In Fig.\,\ref{fig:optical}, we compare the observed optical and
near-UV H and He lines with our final model. Although nebular emission
lines mask the Balmer line centers, we note a Balmer line problem
\citep{1993AcA....43..343N}, that is, the depth of innermost parts of
the observed photospheric profiles of H$\alpha$ and H$\beta$ is not
quite matched by the models. \citet{2010ApJ...720..581G} were unable
to detect this problem because of the lower resolution of their
spectra. In order to fit these lines, a lower temperature would
  be required;  however, this would give too-deep H$\gamma$ and
H$\delta$ line cores and, as mentioned above, would be in conflict
with the metal line strengths. The central emission in \ion{He}{ii}
  $\lambda$4686 is of nebular origin because its height is not
  reached by the emission reversal of the photospheric line profile.

Comparing our results with the most recent ones by \citet[][see
  Sect.\,\ref{sect:intro}]{2010ApJ...720..581G}, the main progress we
achieved is the rather tight constraint of the effective temperature
($\pm 5$\,\% compared to $\pm 22$\,\%) by using ionization balances of
metals. In addition, the error in the surface gravity could be reduced
from about $\pm 0.8$\,dex to $\pm 0.4$\,dex. The same holds for the
He/H abundance ratio. As a consequence, stellar parameters can be
determined with higher precision (Sect.\,\ref{sect:para}).

\subsection{Metal abundances}
\label{sect:metals}

We detected lines from ten metals in the FUSE spectrum
(Figs.\,\ref{fig:egb6_fuse} and \ref{fig:egb6_fuse_detail}). These lines
are well known from previous analyses. Line lists, from which detailed
laboratory wavelengths and transitions can be obtained, were presented by
\citet[][their Tables 2--5]{2015A&A...582A..94W} and \citet[][their
  Tables 2 and 3]{2017A&A...601A...8W}. The derived abundances are
listed in Table~\,\ref{tab:results} and displayed in
Fig.\,\ref{fig:abu}.

From carbon, two \ion{C}{iv} doublets are seen in the FUSE spectrum,
namely the 3p--4d and 3d--4f transitions at 1108\,\AA\ and 1169\,\AA,
respectively. In our final model with C = 0.003, the line cores are
not deep enough to match the observation. A detailed inspection shows
that  the computed line profiles show central emission
cores. This is  unrealistic, but despite some numerical
testing concerning the temperature structure of the model, the reason
remains unknown. Increasing the carbon abundance mitigates the problem
in the line cores, but leads to too strong line wings and to the
appearance of optical \ion{C}{iv} lines, which are not
observed. The variations in \logg, \Teff, and He abundance described
in  Sect.\,\ref{sect:teff} do not improve the \ion{C}{iv}
line fits. The same problem was encountered in our analysis of a hot
DO WD with similar temperature and gravity \citep[PG\,1034$+$001,
  115\,000/7.0; ][their Fig.\,1]{2017A&A...601A...8W}. The star is too
hot to show the \ion{C}{iii} multiplet at 1175\,\AA, which is often
seen in other DAO WDs and in \tol\ (see
Sect.\,\ref{sect:tol26}).

Nitrogen displays the \ion{N}{iv} multiplet at 922--924\,\AA\ and the
singlet at 955\,\AA. We also see two \ion{N}{v} doublets at
1048\,\AA\ and 1050\,\AA, which are partially blended by interstellar
lines. The relative strength of the \ion{N}{iv} and \ion{N}{v} lines
is well matched by our final model. Values of  \Teff\ as high as 115\,000\,K (at
\logg = 7.4) or as low as 95\,000\,K (at \logg = 7.0) can be strictly
excluded because the  \ion{N}{iv} lines are much too weak or too
strong, respectively, while the \ion{N}{v} line strengths are almost
unaffected by the \Teff\ variations.

Prominent oxygen lines from three ionization stages are detected. The
\ion{O}{iv}/\ion{O}{v} line strength ratio is very sensitive against
\Teff\ variations. While the \ion{O}{v} lines hardly change within a
range of 95\,000--110\,000\,K, the \ion{O}{iv} lines quickly disappear
with increasing \Teff. The \ion{O}{vi} $\lambda\lambda$1032/1038
resonance doublet poses a problem. At the adopted abundance, the
innermost line cores are too deep and the line wings are slightly too
shallow. Interestingly, however, two subordinate \ion{O}{vi} lines
(4d--5f and 4f--5g transitions at 1123\,\AA\ and 1125\,\AA) are
reproduced well by our final model. They are too weak in the \Teff =
95\,000\,K model (\logg = 7.0) and too strong in the \Teff =
115\,000\,K model (\logg = 7.4).

A weak \ion{F}{vi} $\lambda$1139.50 line can be detected, allowing for
a fluorine abundance measurement.  The silicon abundance was derived
from a fit to the \ion{Si}{iv} $\lambda\lambda$1122/1128 doublet. The
blue component is blended by one of the mentioned relatively broad
photospheric \ion{O}{vi} lines. Another strong \ion{Si}{iv} doublet at
1067\,\AA\ is blended by a strong interstellar line. The phosphorus
abundance was derived from the \ion{P}{v} $\lambda\lambda$1118/1128
resonance doublet. The sulfur abundance was found from the \ion{S}{vi}
$\lambda\lambda$933/945 resonance doublet, and from two weaker
subordinate \ion{S}{vi} lines at 1000\,\AA\ and 1118\,\AA. The argon
abundance was measured from the rather prominent \ion{Ar}{vii}
$\lambda$1063.55 line, which is blended by the wing of an adjacent
interstellar line. 

The strongest iron lines stem from \ion{Fe}{vii}, followed by
\ion{Fe}{vi} and \ion{Fe}{viii} lines (the only unblended
\ion{Fe}{viii} line is at 1148\,\AA). The relative strengths of
lines from the different ionization stages are sensitive temperature
indicators. For example, at \Teff = 95\,000\,K (and \logg = 7.0) the
\ion{Fe}{vii} lines become too weak, while the \ion{Fe}{vi} lines become
too strong. At \Teff = 115\,000\,K (and \logg = 7.4) the \ion{Fe}{vi}
lines are too weak, while the \ion{Fe}{viii} lines are too strong.  The
nickel abundance is derived from a multitude of relatively weak
\ion{Ni}{vi} lines. 

Other iron group elements were not detected, and we can thus exclude
extreme overabundances that were seen in other DAOs, namely
HS\,2115$+$1148 \citep{2018A&A...609A.107W} and BD$-22^\circ 3467$,
the exciting star of Abell~35 \citep{2012A&A...548A.109Z}. Likewise,
we found no hint of trans-iron group elements that are present in
strongly oversolar amounts in several hot DO WDs \citep[][and
  references therein]{2018A&A...612A..62H}.

To summarize, the abundances of hydrogen, helium, and  ten more
identified species in the white dwarf are, within error limits,
compatible with  solar values. 

\subsection{Stellar parameters, distance}
\label{sect:para}

\begin{figure}[bth]
 \centering  \includegraphics[width=0.9\columnwidth]{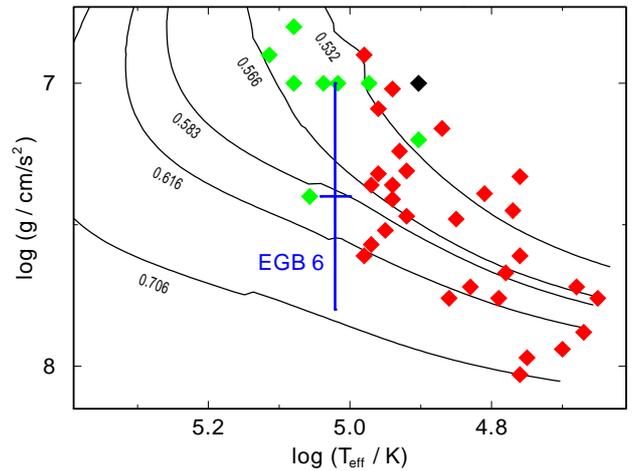}
  \caption{Position of the white dwarf in \egb\ with error bars in the
    $g$--\Teff\ diagram (blue cross). Red, green, and black symbols are
    DAO WDs from the analyses of \citet{2010ApJ...720..581G},
    \citet{2012zieglerdiss}, and \citet{2018A&A...609A.107W}, respectively.
 Black lines are evolutionary tracks of
    post-AGB remnants with masses as labeled \citep[in $M_\odot$,
      from][metallicity $Z=0.01$]{2016A&A...588A..25M}.
} \label{fig:gteff}
\end{figure}

Stellar parameters are estimated from the comparison of the determined
atmospheric parameters \Teff\ and \logg\ with evolutionary tracks by
\citet[][Fig.\,\ref{fig:gteff}]{2016A&A...588A..25M}. We find
$M\,/\,M_\sun=0.58^{+0.12}_{-0.04}$ and
$L\,/\,L_\sun=72^{+300}_{-48}$. The errors are dominated by the
uncertainty in \logg. The post-AGB age is about 4000\,yr; however, the
mass uncertainty causes a large possible error: $\log$($t_{\rm
  evol}$/yr) = $3.60^{+1.26}_{-0.09}$. The mass-loss rates along the
evolution assumed by \citet{2016A&A...588A..25M} imply
$\log$($\dot{M}/M_\sun$/yr) = $-11.0^{+1.1}_{-0.8}$.

The spectroscopic distance $d$ was found by comparing the dereddened
visual magnitude $V_0$ with the respective model atmosphere flux,
resulting in the relation
$$ d {\rm [pc]}= 7.11\times 10^{4} \sqrt{H_\nu\cdot M\cdot 10^{0.4
    V_0-\log g}}\ ,$$
where $H_\nu= 1.60\times 10^{-3}$\,erg cm$^{-2}$s$^{-1}$Hz$^{-1}$ is
the Eddington flux of the model at 5400\,\AA, and $M$ is the stellar
mass in $M_\sun$. For our WD we have V = 15.998
\citep{2016ApJ...826..139B}. From $E(B-V)=0.025$ we derived the visual
extinction using the standard relation $A_V=3.1 E(B-V) = 0.077$,
hence $V_0=15.92$. We found $d=660^{+1100}_{-400}$\,pc. This value is
about 15\% higher than the spectroscopic distance derived by
\citet{2013ApJ...769...32L}, which essentially reflects the ratio of
the effective temperatures of the atmospheric model they used
(93\,235\,K) and that of our model (105\,000\,K). It agrees with the
value of $d=870 \pm 250$\,pc derived by \citet{2016MNRAS.455.1459F}
with a newly developed statistical distance indicator for PNe.
  The distance to the WD given in the Gaia Data Release 2
  ($1.36^{+0.40}_{-0.25}$\,kpc) is a factor of two larger than our result, but in
  agreement within error limits. We conjecture that the
  Gaia parallax measurement may be compromised by the strong
  emission lines from the nearby CEK, which could be variable in
  strength, as was discussed in detail by \citet{2016ApJ...826..139B}.

\begin{table}[t]
\begin{center}
\caption{Parameters for the white dwarf in
  \egb.\tablefootmark{a} }
\label{tab:results} 
\begin{tabular}{cccr}
\hline 
\hline 
\noalign{\smallskip}
\Teff/\,K & $105\,000 \pm 5000$ \\
$\log$($g$/cm/s$^2$) & $7.4 \pm 0.4$      \\
\noalign{\smallskip}
$E(B-V)$  & $0.025\pm0.005$\\
\noalign{\smallskip}
$\log$($n_{\rm H}$/cm$^{-2}$) & $20.3\pm0.4$\\
\noalign{\smallskip}
$M\,/\,M_\sun$ & $0.58^{+0.12}_{-0.04}$ \\
\noalign{\smallskip}
$R\,/\,R_\sun$ & $0.026^{+0.021}_{-0.009}$ \\
\noalign{\smallskip}
$L\,/\,L_\sun$ & $72^{+300}_{-48}$ \\
\noalign{\smallskip}
$d$\,/\,pc      & $660^{+1100}_{-400}$ \\
\noalign{\smallskip}
$\log$($t_{\rm evol}$/yr) & $3.60^{+1.26}_{-0.09}$\\
\noalign{\smallskip}
$\log$($\dot{M}/M_\sun$/yr) & $-11.0^{+1.1}_{-0.8}$\\
\noalign{\smallskip}
\noalign{\smallskip}
\hline 
\noalign{\smallskip}
abundances & $N_i/N_H$            & $X_i$               & [$X_i$]\\
\hline 
\noalign{\smallskip}
H         & 1                    & 0.74                &  0.00 \\
He        & 0.085                & 0.25                &  0.00 \\
C         & $ 3.4 \times 10^{-4}$ & $3.0 \times 10^{-3}$ &  0.10 \\ 
N         & $ 1.9 \times 10^{-5}$ & $2.0 \times 10^{-4}$ & $-0.54$ \\ 
O         & $ 4.3 \times 10^{-4}$ & $5.0 \times 10^{-3}$ & $-0.06$ \\ 
F         & $ 1.1 \times 10^{-7}$ & $1.5 \times 10^{-6}$ &  0.47 \\ 
Si        & $ 3.9 \times 10^{-5}$ & $8.0 \times 10^{-4}$ &  0.08 \\ 
P         & $ 2.2 \times 10^{-7}$ & $5.0 \times 10^{-6}$ & $-0.07$ \\ 
S         & $ 4.3 \times 10^{-6}$ & $1.0 \times 10^{-4}$ & $-0.49$ \\ 
Ar        & $ 2.5 \times 10^{-6}$ & $7.3 \times 10^{-5}$ &  0.00 \\ 
Fe        & $ 8.5 \times 10^{-5}$ & $3.5 \times 10^{-3}$ &  0.46 \\ 
Ni        & $ 2.3 \times 10^{-6}$ & $1.0 \times 10^{-4}$ &  0.17 \\ 
\noalign{\smallskip} \hline
\end{tabular} 
\tablefoot{  \tablefoottext{a}{Mass, radius, luminosity, post-AGB age
    $t_{\rm evol}$, and mass-loss rate $\dot{M}$ from evolutionary
    tracks by \citet{2016A&A...588A..25M}. Abundances are given in number ratios
    relative to hydrogen (column 2), in mass fractions (column 3), and
    logarithmic mass fractions relative to solar value \citep[column
      4; solar abundances from][]{2009ARA&A..47..481A}. Error limits
    for the abundances are $\pm 0.5$\,dex for the metals and $\pm
    0.4$\,dex for helium.}  } 
\end{center}
\end{table}

\subsection{The hot central star of \tol}
\label{sect:tol26}

In Fig.\,\ref{fig:tol26} we show the HST/COS spectrum of
\tol. Photospheric lines from He, C, N, O, and Fe can be identified.

\ion{He}{ii} $\lambda$1640 is dominated by strong nebular emission,
but we  note the outer wings of the photospheric absorption
profile. Shortward of 1180\,\AA, the quality of the spectrum is
somewhat poor but the \ion{C}{iv} doublet at 1169\,\AA\ and the
\ion{C}{iii} multiplet at 1175\,\AA\ are detectable. The \ion{C}{iv}
doublet might be filled in partially by an \ion{He}{i}
$\lambda$1168.67 airglow line (second order from $\lambda$584.33). The
photospheric \ion{C}{iv} resonance doublet is completely filled in by
nebular emission. We note a photospheric \ion{N}{v} resonance doublet,
but a potential photospheric \ion{N}{iv} $\lambda$1718 line is masked
by a tenuous emission feature.  We see a \ion{O}{iv} multiplet at
1340\,\AA\ and \ion{O}{v} $\lambda$1371. Their relative strengths can
be used to constrain the effective temperature. We identify a number
of \ion{Fe}{v} lines, mostly in the 1374--1430\,\AA\ region.

To fit the spectrum, we started to use models with the composition and
surface gravity we adopted for \egb. We adjusted \Teff\ by using the
relative \ion{O}{iv}/\ion{O}{v} line strengths. The best fit is
obtained at \Teff $=85\,000\pm5000$\,K. At that temperature, the
\ion{C}{iii} $\lambda$1175 multiplet (which is not present in
\egb\ because of its higher \Teff) also fits well. In addition, the
\ion{Fe}{v} line features are reproduced, so that the C and Fe
abundances are the same as in \egb. However, we had to reduce the
nitrogen and oxygen abundances. We found N = $5\times10^{-6}$ and O =
$1.5\times10^{-3}$, i.e., 40 and $\approx$\,3 times less than in
\egb. Error limits for abundances are estimated to 0.5\,dex. All other
metal abundances cannot be assessed with the HST spectrum. The
\ion{Si}{iv} resonance doublet is obviously perturbed by interstellar
absorption because the line depths cannot be matched by a photospheric
model even with an unrealistically high Si abundance. The helium
abundance and surface gravity could in principle be constrained with
\ion{He}{ii} $\lambda$1640, but both quantities cannot be derived
independently given  that we only see the outermost part of
the line wings. To check to what extent these uncertainties affect the
determination of the other atmospheric parameters (\Teff, abundances
of C, N, O, Fe), we computed models varying \logg\ ($\pm0.4$\,dex) and
the He abundance (reduction from solar to 1/8 solar) and we found that
our results do not change within error limits, in particular, the N
and O deficiencies remain.

Given the uncertainty in the He abundance, the question arises whether
\tol\ is a DAOZ at all. Could it be a He-dominated  white dwarf (i.e.,
spectral type DOAZ) or even completely devoid of hydrogen (spectral
type DOZ)? Probably not. When we remove hydrogen from our model, the
\ion{He}{ii} $\lambda$1640 wings become so strong and broad that it is
not compatible with the observation, unless the surface gravity is
reduced below \logg = 7 in order to diminish the strength of the line
wings.

\section{Summary and discussion}

\subsection{\egb}

We performed a spectral analysis of the hot WD in the PN
\egb. The effective temperature could be tightly constrained, while the
surface gravity remains relatively uncertain (\Teff = $105\,000 \pm
5000$\,K, \logg = $7.4 \pm 0.4$). A mass of $M/M_\sun =
0.58^{+0.12}_{-0.04}$ was derived. The abundances of H, He, and ten
more species are solar. 

We conclude that element diffusion by gravitational settling and
radiative levitation is not acting in the photosphere. The probable reason is
  that the relatively luminous star ($\approx$\,100\,$L_\sun$)
is losing mass by a radiation-driven wind at a rate on the order of
$10^{-11}\,M_\sun$/yr. Compared to other DAOZ WDs, this is
remarkable because they usually show strong deviations from solar
abundance patterns due to diffusion \citep[e.g.,
][]{2005MNRAS.363..183G,2012A&A...548A.109Z,2018A&A...609A.107W}. The
sample of DAOZs and related hot progenitors analyzed  by
\citet{2012zieglerdiss} contains objects with temperature and gravity
similar to \egb. In particular, the central star of Abell~31 (\Teff =
$114\,000 \pm 10\,000$\,K, \logg = $7.4 \pm 0.3$) has a roughly solar
abundance pattern, while slightly cooler objects like the central star
of Abell~35 (\Teff = $80\,000 \pm 10\,000$\,K, \logg = $7.2 \pm 0.3$)
show extreme deviations from it
\citep{2012A&A...548A.109Z}. Obviously, stellar winds in cooling DAO
WDs effectively prevent element diffusion as long as \Teff\ is higher
than $\approx$\,100\,000\,K. Our result for \tol\ (\Teff
$\approx$\,85\,000\,K; O and N depletion) confirms this
conclusion. Similarly, the correlation between helium abundance and
effective temperature found by \citet{2010ApJ...720..581G} in their
sample of DAOs was interpreted as an indication of how the weakening
stellar wind along the WD cooling sequence can no longer maintain
helium in the atmosphere of a white dwarf.

The theoretical mass-loss rate from the hot WD,
$\log$($\dot{M}/M_\sun$/yr) = $-11.0^{+1.1}_{-0.8}$, is  orders of
magnitudes larger than required to inhibit Bondi--Hoyle accretion of
interstellar material \citep{1992ApJ...394..619M}. We therefore do not
expect that any circumstellar material from the PN is accreted unless
the accretion flow is reorganized into a disk, due to angular
momentum. 

Concerning the outer nebula, \citet{1989ApJ...346..251L} pointed out a
discrepancy between their estimations of the post-AGB age of the WD
\citep[several $10^5$\,yr; based on their spectral analysis and
  evolutionary tracks by][]{1986A&A...154..125K} and the kinematical age
of the nebula (20--50\,kyr, based on their derived spectroscopic
distance of 460\,pc and assuming a PN expansion rate of
20--50\,km/s). Our results indicate a much younger WD age,
$\log$($t_{\rm evol}$/yr = $3.60^{+1.26}_{-0.09}$), mainly due to the
higher effective temperature (105\,000\,K compared to
70\,000\,K). Taking the PN angular extension ($13\arcmin$), the
spectroscopic distance derived by us (660\,pc), and the expansion
velocity \citep[38\,km/s,][]{1990A&A...232..129H}, we obtain a
kinematical age of 15\,kyr, which is  in agreement with the post-AGB age of the
WD. 

If the material around the companion was captured at the PN ejection
15\,kyr ago, and if the disk lifetime is not much longer, then this
would explain why the \egb\ phenomenon is uncommon. Also, the large
distance of the companion from the primary (at least
$\approx$\,100\,AU) prevents a fast evaporation of the disk material
by UV radiation from the primary. Measured orbital periods of binary
central stars are all much shorter (from hours up to
$\approx$\,10\,yr; see, e.g., the compilation maintained by
D\@. Jones\footnote{\url{http://drdjones.net/bCSPN}}) than that of
\egb\ (larger than $\approx$\,500\,yr); however a number of wide,
visual binaries are known as well \citep[separations of $\approx$\,100
  to a few thousand AU,][]{1999AJ....118..488C}.

\citet{2013ApJ...769...32L} considered an alternative explanation for
the origin of the compact emission knot. They put forward the idea
that it is the region where the winds from the WD and the supposed dM
companion collide. Their ``duelling-winds'' model  was considered
unlikely by \citet{2016ApJ...826..139B} since they discovered that the
companion is not an exposed dM star but a dust-enshrouded
low-luminosity star. In view of our result that the WD is rather young
and luminous, it might be worth reconsidering the colliding-wind
hypothesis. From an ionization model for the CEK,
\citet{2013ApJ...769...32L} estimated that a WD wind with a rather
high mass-loss rate on the order of $10^{-9}$\,M$_\sun$/yr would be
required, but they questioned how a hydrogen-rich WD could maintain
such a high rate.

We have shown that the WD has a solar metallicity; therefore, a
radiation-driven wind can be sustained with a mass-loss rate of up to
$10^{-10}$\,M$_\sun$/yr, and thus almost reaches the required
value. However,  we also point out that there is no spectroscopic hint of
mass-loss; in particular, we would expect the \ion{O}{vi} resonance
doublet to be most susceptible to wind effects, but the line centers
are not shifted relative to other photospheric lines and the profiles
are symmetric.  Investigations of H-rich central stars showed that the
rates are usually larger than $10^{-9}$\,M$_\sun$/yr. The lowest rate
derived so far, and thus constraining the threshold for the
spectroscopic proof of existing winds, is that of NGC\,1360 with
$\approx 10^{-10}$\,M$_\sun$/yr \citep{2011MNRAS.417.2440H} using
very weak deviations of the \ion{O}{vi} resonance doublet line
profiles from hydrostatic-atmosphere profiles.  Detailed expanding
model-atmosphere computations tailored to \egb\ are required to determine the mass-loss rate that would be necessary to leave a detectable
spectroscopic signature.

\subsection{\tol}

We analyzed a UV spectrum of \tol. We found, as for \egb, that the
hot nucleus of \tol\ is a DAOZ WD, but cooler (\Teff
$\approx$\,85\,000\,K). Helium abundance and surface gravity are
uncertain with the observations currently at hand. We measured the
abundances of C, N, O, and Fe. Significant depletion of nitrogen and
oxygen was revealed, which can be interpreted as an indication that
the gravitational settling of elements affects the abundance pattern in
the photosphere, in contrast to the WD in \egb. Our results confirm
that, with regard to the hot nucleus and the compact nebula emission region,
\tol\ is similar to \egb. It remains to be investigated whether the
central star of \tol\ also has a low-luminosity companion.

\begin{acknowledgements} 
We thank R\@. Napiwotzki for putting the SPY survey spectra at our
disposal. The TMAD tool (\url{http://astro.uni-tuebingen.de/~TMAD})
used for this paper was constructed as part of the activities of the
German Astrophysical Virtual Observatory. Some of the data presented
in this paper were obtained from the Mikulski Archive for Space
Telescopes (MAST). STScI is operated by the Association of
Universities for Research in Astronomy, Inc., under NASA contract
NAS5-26555. Support for MAST for non-HST data was provided by the NASA
Office of Space Science via grant NNX09AF08G and by other grants and
contracts. This research has made use of NASA's Astrophysics Data
System and the SIMBAD database, operated at CDS, Strasbourg,
France. Funding for SDSS-III has been provided by the Alfred P. Sloan
Foundation, the Participating Institutions, the National Science
Foundation, and the U.S. Department of Energy Office of Science. This
work has made use of data from the European Space Agency (ESA) mission
{\it Gaia} (\url{https://www.cosmos.esa.int/gaia}), processed by the
{\it Gaia} Data Processing and Analysis Consortium (DPAC,
\url{https://www.cosmos.esa.int/web/gaia/dpac/consortium}). Funding
for the DPAC has been provided by national institutions, in particular
the institutions participating in the {\it Gaia} Multilateral
Agreement.

\end{acknowledgements}

\bibliographystyle{aa}
\bibliography{aa}

\end{document}